\documentclass[journal,comsoc]{IEEEtran}

\usepackage{caption}
\usepackage[T1]{fontenc}

\ifCLASSINFOpdf
            \else
                  \fi

\usepackage{amsmath}
\usepackage[cmintegrals]{newtxmath}

\usepackage{graphicx}
\graphicspath{{figures/}}
\DeclareGraphicsExtensions{.jpg,.png}
\usepackage{etoolbox}
\usepackage[caption=false,font=footnotesize]{subfig}
\usepackage{floatrow}
\newfloatcommand{capbtabbox}{table}[][\FBwidth]

\usepackage{multirow,makecell}\usepackage[table]{xcolor}
\usepackage{amssymb}
\usepackage{amsmath}
\usepackage{booktabs}

\usepackage{framed}
\usepackage{float}

\usepackage{listings}

\usepackage{array}
\usepackage{rotating}
\usepackage{adjustbox}
\usepackage{tabularx}
\usepackage[framemethod=default]{mdframed}
\usepackage{tikz}
\usepackage{pgfplots, pgfplotstable}
\usetikzlibrary{patterns}

\usepackage{pgfplotstable}
\usepackage{filecontents}

\usepackage{array}
\usepackage{xcolor}
\usepackage[hyphens]{url}
\usepackage[pdftitle={A Framework and Comparative Analysis of Control Plane Security of SDN and Conventional Networks},pdfauthor={Abdou},unicode,hidelinks]{hyperref}
\usepackage{url}
\usepackage{adjustbox}
\usepackage{fancyvrb}
\usepackage{balance}
\usepackage{xspace}
\newcommand{\etal}{\textit{et al.}\xspace}

\newcommand{\ie}{\textit{i.e.,}\xspace}
\newcommand{\eg}{\textit{e.g.,}\xspace}
\newcommand{\cf}{\textit{cf.}\xspace}

\newcommand\blfootnote[1]{  \begingroup
  \renewcommand\thefootnote{}\footnote{#1}  \addtocounter{footnote}{-1}
  \endgroup
}

\newcounter{protocol}[section]

\newcolumntype{T}[2]{    >{\adjustbox{angle=#1,lap=\width-(#2)}\bgroup}
l    <{\egroup}}

\newcolumntype{L}[1]{>{\raggedright\let\newline\\\arraybackslash\hspace{0pt}}m{#1}}
\newcolumntype{C}[1]{>{\centering\let\newline\\\arraybackslash\hspace{0pt}}m{#1}}
\newcolumntype{R}[1]{>{\raggedleft\let\newline\\\arraybackslash\hspace{0pt}}m{#1}}

\newcommand{\ltwo}{L2}
\newcommand{\lthree}{L3}
\newcommand{\CN}{CN}

\newcommand{\edgeonly}{\textsc{end-host}}
\newcommand{\allelement}{\textsc{all-element}}

\pgfplotsset{compat=1.8}

\begin{document}
\title{A Framework and Comparative Analysis of Control Plane Security of
SDN and Conventional Networks}

\author{

\IEEEauthorblockN{\hspace{-0.8cm} AbdelRahman Abdou and Paul C. van Oorschot \hspace{3cm} Tao Wan}

\IEEEauthorblockA{\hspace{1cm} {Carleton University, Canada} \hspace{4.4cm} {Huawei Canada}}

}

\maketitle

\begin{abstract}

Software defined networking implements the network control plane in an
external entity, rather than in each individual device as in conventional
networks. This architectural difference implies a different design for control
functions necessary for essential network properties, \eg loop prevention
and link redundancy. We explore how such differences redefine the security
weaknesses in the SDN control plane and
provide a framework for comparative analysis which focuses on essential
network properties required by typical production networks. This enables
analysis of how these properties are delivered by the control planes of SDN
and conventional networks, and to compare security risks and mitigations.
Despite the architectural difference, we find similar, but not identical,
exposures in control plane security if both network paradigms provide the
same network properties and are analyzed under the same threat model. However,
defenses vary; SDN cannot depend on edge based filtering to protect its control
plane, while this is arguably the primary defense in conventional networks. Our
concrete security analysis suggests that a distributed SDN architecture that
supports fault tolerance and consistency checks is important for SDN control
plane security. Our analysis methodology may be of independent interest for
future security analysis of SDN and conventional networks.
\end{abstract}


\IEEEpeerreviewmaketitle

\vspace{-0.1cm}
\section{Introduction}
\vspace{-0.1cm}
\blfootnote{Version: \today}

Software-Defined Networking is a relatively new network architecture in
which the control plane is separated from each individual network device
and instead implemented in an external software entity.  The external entity
has complete knowledge of the topology of a network under its control, and
programs the forwarding tables of each individual device in the network. In
contrast in \emph{conventional networks} (\CN{}s),
the control plane, including implementation of \eg a routing protocol such
as Open Shortest Path First (OSPF)~\cite{rfc2328}, runs inside each network
device to learn forwarding tables in a distributed fashion.
SDN architectures have two distinguishing properties of direct interest
herein~\cite{feamster2014road}:

\begin{enumerate}
    \item
{\bf Control and data plane separation}.
Removing the control plane from network devices and implementing it in
an external SDN controller significantly reduces the complexity of network
devices, making them simpler and cheaper than \CN{} devices whose distributed
control plane functionality is implemented across millions of lines of code,
defined across hundreds of RFCs.
\item
{\bf Network programmability}.
An SDN controller, with complete knowledge of a network's topology, controls
a multitude of network devices within its administrative domain. By providing
application programming interfaces (APIs), SDN makes it possible to develop
networking applications, \eg traffic engineering \cite{rfc7471}, thus enabling
network innovation. In contrast, \CN{} devices are proprietary and closed,
making it hard or impossible to develop innovative network applications.

\end{enumerate}

The concept of SDN has evolved since the term was originally coined in
2009~\cite{greene2009MIT}. Here we try to clarify the critical properties of
SDN from the perspective of network devices.  A network device can be pure
SDN, non-SDN, or hybrid. A pure SDN device implements no control function
and is fully controlled by an external SDN controller. A non-SDN device
implements all of its own control functions and is not controlled by any
SDN controller. A hybrid SDN device both implements control functions,
and is controlled by an SDN controller. Based on this classification of
network devices, a network can also be one of three types. A pure SDN
network consists of at least one SDN controller and network devices all of
which are fully controlled by the controller. A non-SDN network (\ie \CN{})
consists of network devices all of which implement and run their own control
functions with no controlling external entity. A hybrid network consists of
hybrid devices and at least one SDN controller.

In academic work, ``SDN" often implies a pure SDN network, such as an OpenFlow
network, and many academic SDN security research papers (\eg~\cite{SHI13})
focus primarily on the security of OpenFlow networks. SDN controllers
originating in academic work, such as FloodLight and NOX, also primarily
support OpenFlow and control OpenFlow switches which implement no control
functionality (\ie are pure SDN, rather than hybrid).

In contrast in industry, SDN commonly refers to hybrid networks consisting
primarily of \CN{} devices, augmented with open interfaces also allowing
external control by an SDN controller. For example Broadcom, a leading
provider of switch chips, published OpenFlow Data Plane Abstraction
(OF-DPA) software~\cite{broadcomOFDPA} to allow switches based on Broadcom
chips to be controlled by OpenFlow. Note those \CN{} devices, although
often claimed to support OpenFlow and which can be controlled using the
OpenFlow protocol, do not actually implement OpenFlow tables and are
not true OpenFlow switches. Rather, they use conventional tables such as
\lthree{} tables and Access Control Lists (ACLs) to emulate the behavior of
OpenFlow tables, which allows packets to be processed beyond destination
addresses. As another example, OpenDayLight~\cite{medved2014opendaylight}
and ONOS~\cite{berde2014onos}, two leading open source SDN controllers, can
control not only OpenFlow switches but also conventional devices, \eg using
NETCONF~\cite{rfc6241}. It is clear that industrial network practitioners
focus more on network programmability than on the separation of control and
data planes. We refrain from speculating on which type of SDN is better,
or is the future.

We study and compare the control plane security of a pure SDN (hereafter
referred to as SDN) and a \CN{}. While hybrid networks are more popular in
the field, there is no clear consensus on how to best divide control functions
locally inside a device and externally into a controller. Further, by studying
the security of both SDN and \CN{}s, we hope that security threats identified
in each  can be selectively applied to a given hybrid network when its local
and external controls are well defined.

Research on the security of SDN and \CN{}s is in two distinct states. On one
hand, the security of \CN{}s has received less academic attention but is well
understood by network security practitioners; aside from the area of routing
(\eg BGP security~\cite{van2007interdomain}) there are relatively few academic
papers on the control plane security of a \CN{}, security threats are well
understood by equipment vendors and many security mitigations are built into
\CN{} products (\eg switches, routers). In contrast, SDN security has received
considerable academic attention (\eg \cite{HONG_NDSS15,POR_NDSS15,DHA_NDSS15}),
but its progress is considered slow (at best) by industrial measures. For
example, neither of the two leading open source SDN controllers, OpenDaylight
and ONOS, has implemented significant security mitigation.

These different states of SDN and \CN{} security research have attracted little
attention. We offer the following explanation. We observe that many papers on
SDN security assume a simple network, ignoring practical properties such as
redundancy and scalability essential to realistic networks---thus excluding
security risks faced by important network control functionality. Further,
security threats identified for SDN are not properly compared with those
in \CN{}. We suggest that the lack of academic scrutiny, particularly
systematization literature, on \CN{} security, may be a significant
contributing factor in the considerable academic research on SDN security
having failed to have major impact in industry.

We aim to address the gap by a comparative security assessment of conventional
and SDN networks. Rather than a security analysis of all aspects, we focus
on control plane security, since it is in their control plane architecture
that \CN{}s and SDN differ  primarily.

We provide a framework consisting of essential network properties required by
production networks. Using this, we study how those properties are achieved by
SDN and \CN{} respectively, and analyze the security risks and mitigations
accordingly. Our finding is that the security threats faced by SDN and
\CN{} are comparable in an apples-to-apples comparison, \ie if despite
the architectural differences between the two types of networks, they are
tasked to provide the same network properties under the same threat model.
However, defenses vary in that filtering in the network edge is effective in
\CN{}, but less so in SDN. Further, consistency checks, which are required
by both networks to defeat inside attacks, can be implemented inside each
\CN{} device, but require a highly modularized SDN software architecture
to facilitate implementation there. Our finding is supported by detailed
security analysis. We argue that our methodology for comparative analysis
will be of independent interest, to guide future SDN security analysis in
both academia and by  practitioners.

The sequel is organized as follows. Section~\ref{sec:background}
provides background information on \CN{} and SDN
architecture. Section~\ref{sec:framework} outlines fundamental network
properties required by typical production networks, as well as the threat model
used for our analysis. Sections~\ref{sec:layer2} and~\ref{sec:layer3} analyze
the security risks of the control plane of conventional Layer-2 (\ltwo{})
and Layer-3 (\lthree{}) networks respectively.  Section~\ref{sec:sdn} analyzes
security risks in SDN networks. We compare the security risks and mitigation
of SDN with \CN{} in Section \ref{sec:comparison}.  Section~\ref{sec:related}
reviews related work.  Section \ref{sec:discussion} concludes.

\section{Background}
\label{sec:background}

Here we provide background on conventional and SDN networks for consistent
terminology and later reference.  Networking experts may advance to Section
\ref{sec:framework}.
\subsection{Conventional Networks}

A \CN{} can be \ltwo{} or \lthree{}.  A network consisting of only \ltwo{}
switches as its intermediate systems is called an \ltwo{} network. Two (or
more) \ltwo{} networks can be connected, \eg using an \lthree{} router. A
network of \lthree{} routers is called an \lthree{} network. Other than
using different types of destination addresses for forwarding, \ltwo{}
and \lthree{} networks differ mainly in two aspects:

\begin{enumerate}
    \item
      They use different mechanisms in constructing their forwarding tables.
      \ltwo{} devices learn their forwarding tables (\ie MAC tables) from the
      data plane.  \lthree{} routers build routing tables from the control
      plane using routing protocols. Note: MAC tables map MAC addresses to
      switch ports, not to be confused with ARP tables which map IP addresses
      to MAC addresses.
    \item

      They handle unknown packets differently. An unknown packet is a packet
      without any corresponding forwarding rules.  An \ltwo{} device floods
      an unknown packet to all ports except the receiving one to learn the
      forwarding rule, while an \lthree{} router drops an unknown packet
      (and may also notify the packet source, \eg using ICMP).

\end{enumerate}
Due to these differences, \ltwo{} and \lthree{} networks face different
sets of security risks. Thus, we divide \CN{} into \ltwo{} and \lthree{},
and discuss separately in Sections~\ref{sec:layer2} and \ref{sec:layer3}.

\subsection{Software Defined Networking}

\begin{figure}[]  \centering
  \includegraphics[scale=0.4]{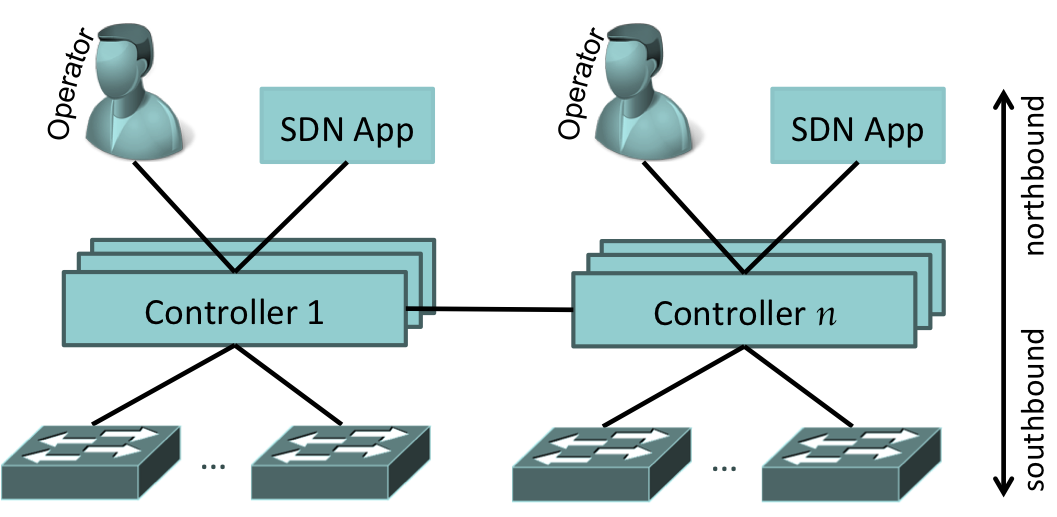}
  \caption{Generic SDN architecture}
  \label{fig:sdn-arch}
\end{figure}

SDN involves one or more SDN controllers, each controlling a number
of network elements within its domain via standard protocols (such as
OpenFlow). Each controller may run in multiple instances, each further
managing a subset of network elements and backing-up other instances to
provide both scalability and high availability. SDN controller instances also
communicate with each other within the same domain, or may be federated with
controllers in other domains, \eg to form a complete view of the network
(see Fig.~\ref{fig:sdn-arch}). Further, there may be a hierarchy of SDN
controllers for scalability or multiple layer control.

An SDN controller usually provides a web based portal and APIs for network
operators and applications to control network elements respectively. The
interfaces between SDN controller and network elements are labelled {\it
southbound}, and those between SDN controller and SDN applications are
{\it northbound}.

\subsubsection*{Scope of our Analysis}

An SDN controller is an entity that does not exist in a \CN{}, thus
its security requires special attention. As noted earlier, we focus
on control plane security herein; this area has not received much
attention~\cite{HONG_NDSS15}, motivating us to consider it and herein give
a framework for directly comparing control plane security issues with those
facing \CN{}s. We see analysis of control plane security as an important
step contributing to a broad security analysis of SDN, which should include
all SDN components (see Fig.\ref{fig:sdn-arch}).

\section{Framework}
\label{sec:framework}

Our framework consists of a set of properties required by typical production
networks, and two threat models (see below) to be applied to both SDN
and \CN{}s.

\subsection{Network Properties}

Production networks must provide properties such as loop free forwarding
to allow entities attached to the network to communicate.
Here we outline five primary such properties. While these are not specifically
related to security, they are important in security analysis as each may
require its own control functions and introduce unique security risks. Since
multiple properties may be provided by a common control protocol, each
property does not necessarily introduce new security risks.

\begin{itemize}
    \item{\textbf{{Basic Forwarding}}. A network consisting of a single
    switch must establish forwarding information to allow   attached entities
    to communicate.}
    \item{\textbf{Loop Free Forwarding}. A network consisting of multiple
    devices and links which form physical loops must ensure there is no
    forwarding loop among network device forwarding tables.}

    \item{\textbf{Link Redundancy}. If there are multiple links between a
    pair of network devices, the network topology should remain unchanged in
    the event that one or several of such links go down as long as there is
    one functioning link between the pair. Further,  it should be possible
    to use all links to transmit data, instead of only one.  }

    \item{\textbf{Device Redundancy}. This property is often referred to as
    \textit{high availability}. A network consisting of two or more devices
    should remain fully available in the event of the failure of any single
    device.}
    \item{\textbf{Scalability}. As a network grows and becomes large, it
    should remain functional and manageable. Network  design should allow
    growth without significant management overhead. }

\end{itemize}

Besides these properties, networks typically must provide other
properties such as QoS and certain services to address user needs. For
example, a carrier network may need to support virtual private networks
(VPN)~\cite{sharafat2011mpls} to its customers, \eg using MPLS. Control
protocols required to provide those network services usually introduce their
own security risks. However, discussion of such control protocols is beyond
our present scope.

\subsection{Threat Models}
\label{sec:trustmodel}

We consider two threat models: \edgeonly{}, where only end hosts are assumed
vulnerable to compromise or under attacker control, and \allelement{}, where
all network elements are assumed vulnerable. Under the \edgeonly{} threat
model, SDN controllers and the network devices themselves are considered
trustworthy but end hosts attached to the network are not. In contrast,
neither SDN controllers nor network devices are considered trustworthy under
the \allelement{} threat model; in other words, attacks could be from hosts,
as well as network devices and SDN controllers. We use these two threat
models in Section~\ref{sec:comparison} to compare the security risks and
defenses of SDN and \CN{}.

\section{\ltwo{} Networks}
\label{sec:layer2}

We now discuss how \ltwo{} networks satisfy the network properties of our
framework, and analyze security risks and mitigations associated with each
property.

\subsection{Basic Forwarding}
\label{subsec:Connectivity}

\begin{figure}[h]
\mbox{
\subfloat
{
\vspace{-100pt}
\includegraphics[scale=0.4]{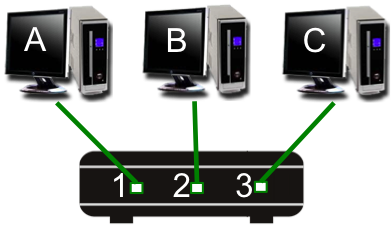}
}
\hspace{10pt}
\subfloat
{
\scalebox{0.8}{
\begin{tabular}{r|l}
Port \# & MAC \\
\hline 
1 &  A \\
2 & B \\
3 & C \\
\multicolumn{2}{c}{}\\
\multicolumn{2}{c}{}\\
\multicolumn{2}{c}{}\\
\multicolumn{2}{c}{}\\
\end{tabular}
}
\vspace{100pt}
}
\caption{MAC table for switch with three devices connected.}
\label{fig:lay2-1switch}
}
\vspace{-35pt}
\end{figure}

To provide network connectivity, an \ltwo{} device uses MAC learning to
build its MAC table, which maps switch ports to MAC addresses, and possibly
other information such as VLAN tags. When a switch receives a frame from
one of its physical ports, it adds a new (or updates an existing) entry
in the MAC table mapping the frame's source MAC address to the receiving
port. Multiple MAC addresses can be associated with a same port number. To
forward a frame, a switch looks up its destination MAC address in the table,
and forwards the frame through the corresponding port. If no entry is found,
the switch floods the frame to all ports except the receiving one. The
intended destination, upon receiving the flooded frame, sends a response
frame enabling the switch to learn the mapping between receiving port and
responding source MAC address. The missing mapping entry is added into the
table for future use. Figure~\ref{fig:lay2-1switch} illustrates a MAC table
for a single switch.

\subsubsection{Attacks}
\label{subsubsec:Attacks}

An \ltwo{} MAC table is learned from data plane (including end-user)
packets. Thus, it is subject to {\it MAC attacks}~\cite{convery2002hacking}. A
malicious host can send a packet with a falsified source MAC address to
poison a switch's MAC table.   Two known attack strategies are as follows. In
\textit{MAC spoofing}, an attacker sends frames with spoofed source MAC
addresses matching those of target (victim) hosts, thereby hijacking traffic
destined to those victims. If a victim is actively sending packets to switches,
the poisoned MAC table will   alternate between correct and falsified states. A
more effective attack, \textit{MAC Flooding}, sends a large number of garbage
frames with randomly generated source and destination MAC addresses to fill
up the MAC table. Once the table is full with non-existent MAC addresses,
legitimate frames will not match any forwarding entry, resulting in flooding
of frames to switch ports including those connecting to the attacker who can
thus eavesdrop or even hijack virtually all traffic. Any device, including
end-user devices (outsiders) and network devices (insiders), can similarly
manipulate a MAC table.

Note that this differs from ARP spoofing (Section
\ref{subsubsec:layer3basicAttacks}). MAC attacks poison the MAC table of
a switch using any packet with a spoofed source MAC address. ARP spoofing
poisons the ARP table of a host or router using only ARP-related packets
(\eg ARP response or gratuitous ARP \cite{fall2011tcp}).

\subsubsection{Defenses}

MAC attacks can be mitigated by preventing untrusted devices such as hosts
from sending packets with spoofed MAC addresses. One such mitigation mechanism,
{\it port security}~\cite{ciscoportsec}, allows a switch to bind a port to one
or several MAC addresses (\textit{MAC binding}). Port security usually also
allows limits on the number of MAC addresses to be associated with a switch
port (\textit{MAC limiting}). MAC binding can prevent MAC spoofing, but is
static and typically requires manual configuration---possibly introducing
configuration overhead and misconfigurations. MAC limiting appears more
practical as it can mitigate MAC flooding attacks and requires only simple
configuration, but alone, does not prevent MAC spoofing. 

\subsection{Loop Free Forwarding}

\begin{figure}[ht]
\centering 
\includegraphics[scale=0.5]{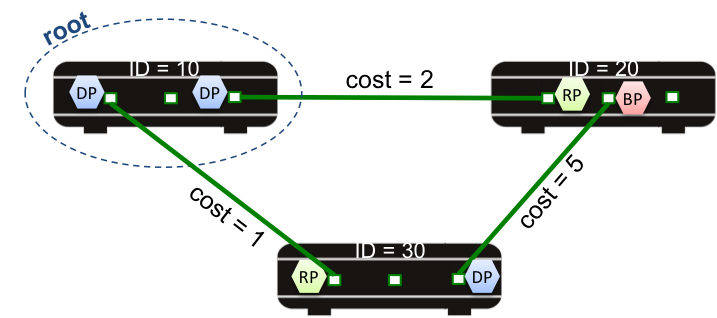}
\caption{Operation of the spanning tree protocol.}
\label{fig:layer2-3switch}
\end{figure}

In \ltwo{} networks with multiple switches and looping links (see
Fig.~\ref{fig:layer2-3switch}), forwarding loops can occur when unknown
\ltwo{} frames are flooded and no Ethernet frame TTL field limits how many
times a frame is forwarded. Thus for loop prevention, an \ltwo{} control
protocol like the Spanning Tree Protocol (STP)~\cite{perlman1985algorithm}
is needed. Switch ports connecting other switches and end user hosts are
often referred to as Network-to-Network Interfaces (NNIs) and User-to-Network
Interfaces (UNIs) respectively.

In STP, switches exchange Bridge Protocol Data Units (BPDUs) carrying
information about switch identifiers and path costs, and accordingly compute
a spanning tree. A root switch is first elected, typically that with the
smallest identifier. Each non-root switch then determines the {\it root port}
as the port  with least-cost path leading to the root switch. Similarly,
for each link in the network, the end port closer to the least-cost path
is called the {\it designated port}. All remaining ports are called {\it
blocked ports}. A spanning tree then consists of all the network switches
(one as  root) and some network links. The links not in the spanning tree
are still used to exchange control plane traffic (\eg BPDU) but only in one
direction to be loop-free.
\subsubsection{Attacks}

STP uses BPDU, with a multicast destination MAC address, to exchange
topology information to elect a root bridge and to establish a spanning tree,
assuming all BPDUs are trustworthy. Due to the lack of security protection
(\eg no default robust authentication \cite{attack_STP}), STP is subject
to BPDU spoofing attacks~\cite{kiravuo2013survey}, as well as \textit{BDPU
tampering} and \textit{BDPU flooding}. For example, an attacker could send
a spoofed BPDU packet with a low priority and small MAC address to result
in the lowest bridge identifier among all switches, thus winning the root
bridge election. Being the root bridge, the attacker receives virtually
all network traffic within the STP domain. BDPU tampering could lead to the
calculation of incorrect network topology. BDPU flooding could force switches
to continuously re-calculate  topology,  resulting in service disruption. 

\subsubsection{Defenses}

An administrator may intervene in root placement, \eg manually specifying
the location of the root switch; Cisco's {\it root guard} command
\cite{ciscosecindepth} facilitates this. Likewise, {\it BPDU filter} prevents
a host from participating in STP by filtering BPDUs in NNIs. 

\subsection{Link Redundancy}

While STP can prevent forwarding loops, it uses a single link between a pair of
switches even when redundant links exist, resulting in underutilized network
bandwidth or even packet loss in the event of link failures. To improve
bandwidth usage and redundancy, \ltwo{} protocols may support \emph{link
aggregation}, grouping multiple links into one virtual link. A Link Aggregation
Group (or LAG), viewed as a single link, can be included in a spanning tree,
allowing their collective use for link protection and load balancing.

Link aggregation can be configured manually, or established dynamically by
the Link Aggregation Control Protocol (LACP)~\cite{ieee_802_1AX_2014}. LACP
transmits LACP Data Unit (LACPDU) to inform the other end ({\it partner})
of its state and its understanding of partner state. Based on LACPDU, a LAG
can be dynamically created and updated.

\subsubsection{Attacks}

A switch running LACP sends to, and receives from its partner, LACPDUs to
maintain link aggregation.  LACPDUs are typically sent over a point-to-point
link, making man-in-the-middle tampering difficult, but remaining vulnerable
to \textit{LACP spoofing} attacks because (1) LACP is usually implemented in
the CPU (vs.\ data plane), allowing a switch to receive LACPDUs from remote
entities; and (2) it has no security protection (\eg peer authentication)
\cite{ieee_802_1AX_2014}. Thus, an external entity (\eg a host) may send
forged LACPDUs to a switch to influence the state of its link aggregation,
\eg to cause link instability or even denial of service. 

\subsubsection{Defenses}

Implementing LACP in the data plane may ensure that LACPDUs are only
received from a given port and never leave that port, mitigating forged
LACPDU injection.

\subsection{Device Redundancy}

Device level redundancy ensures that the failure of one or more devices
does not result in loss of network connectivity; \eg  Cisco Switch Stacking
(CSS)~\cite{ciscoCSS} allows a number of switches, usually of the same model,
to form a redundancy group. Within a redundancy group, a master is elected
dynamically, \eg based on bridge identifiers and/or priority values. If a
master fails, a new master is elected to ensure ongoing network connectivity. 

\subsubsection{Attacks}
Electing a switch master by dynamic election is subject to election spoofing
attacks. An adversary may send falsified messages to become the master. 

\subsubsection{Defenses}
The election process should use cryptographic methods for origin authentication
and message integrity to exclude unauthorized entities, \eg a host, from
joining. 

Switches within a redundancy group should be connected via dedicated ports,
and the election process should only run in those ports. Any election message
received from other ports should be dropped. 

\subsection{Scalability}

Scalability is an important issue with STP and MAC learning; every switch
must learn all MAC addresses and identifiers in the network. While easy
in small networks, challenges arise in larger networks such as a large
enterprise data center with many physical servers and virtual machines,
each with several MAC addresses. The number of MAC addresses may exceed the
MAC table capacity of a switch, and increase the delay of MAC table look ups.

To improve scalability, VLANs may be used to divide a network into segments,
each an isolated \ltwo{} broadcast domain. MAC learning then occurs within
a VLAN, reducing MAC table size. \lthree{} routing is used to connect VLANs.

\begin{figure}[!ht]
  \centering {
  \includegraphics[scale=0.5]{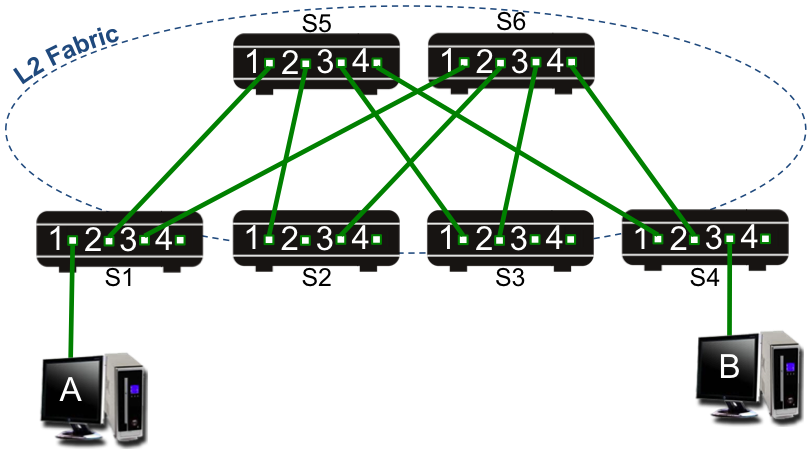}
  \caption{\ltwo{} network with network fabric}
  
  \label{fig:lay2-fabric}
  }
\end{figure}

Another method to improve scalability is to group core switches into domains
called {\it network fabric}. In Fig.~\ref{fig:lay2-fabric}, switches S5,
S6 and ports on switches S1--S4 (customer edge switches) form a fabric and
run an \ltwo{} routing protocol, \eg Intermediate System to Intermediate
System (IS-IS) \cite{rfc5304}, to learn the fabric network topology. In
Fig.~\ref{fig:lay2-fabric}, a frame from A to B is encapsulated with an
outer header with S1 as source address and S4 as destination address. The
outer header usually introduces TTL to prevent indefinite forwarding within
the fabric. Each customer edge switch learns all local MAC addresses and
some remote MAC addresses; S5 and S6 learn no end-user MAC address. Example
network fabric implementations are Transparent Interconnection of Lots of
Links (TRILL)~\cite{rfc6325} and Shortest Path Bridging (SPB)---see IEEE
802.1aq~\cite{ieee_802_1aq}.

\subsubsection{Attacks}

VLANs are subject to \textit{VLAN hopping attacks}---traffic from one VLAN
can be received by another, allowing \ltwo{} attacks against one VLAN to be
launched from a different VLAN. One attack strategy is \textit{VLAN double
encapsulation}~\cite{VLAN_attacks}. Another attack strategy is to exploit
switch misconfiguration or VLAN auto negotiation protocols to impersonate
another switch. In this way, a malicious host can pretend to be a switch
and the link between the host and a switch would appear to be a trunk link,
allowing the host to send and receive packets with any VLAN tag. Thus,
the network isolation provided by VLAN is completely broken.

There are multiple means for implementing \ltwo{} network fabric, \eg TRILL and
SPB, all of which use a routing protocol, typically IS-IS, to automatically
discover and maintain the network topology inside the fabric. IS-IS is
subject to several attacks, such as PDU spoofing, and DoS due to replaying
hello messages \cite{rfc7176,rfc6325}.

\subsubsection{Defenses}

VLAN-related vulnerabilities may be mitigated by disabling VLAN
auto-negotiation, and configuring VLAN filtering in UNIs. Note that a packet
may contain more than one VLAN tag; all such tags should be filtered if
present in packets received from UNIs. 

IS-IS vulnerabilities may be mitigated by enabling additional IS-IS
cryptographic authentication~\cite{rfc5304}, and ignoring unauthenticated
PDUs. Further, IS-IS messages received from UNIs should be dropped. 

\section{\lthree{} Networks}
\label{sec:layer3}

\subsection{Basic Forwarding}
\label{subsec:layer3basic}

\lthree{} devices, namely routers, perform two main tasks: route learning
and packet forwarding \cite{book:kurose}. In simple networks where two or
more subnets are connected by a single router, the routing table is usually
manually configured without running any routing protocol. A packet sent
from one subnet to another has a destination MAC address of the default
gateway of the source host, thus always arrives at the router via one of its
interfaces. A router, upon receiving a packet from one subnet destined to
another, performs three actions: removing the packet's L2 header, looking up
the routing table for the next hop, and encapsulating the packet with a new
L2 header for forwarding. A next hop in a routing table could be a local
interface or an IP address. Routing table looking up is recursive until
a next hop is a local interface. In this case, it further looks up the ARP
table associated with that interface for the MAC address of the packet's next
hop IP address. If not found, the router uses ARP to obtain the MAC address.

\subsubsection{Attacks}
\label{subsubsec:layer3basicAttacks}
Since an \lthree{} router uses ARP  to resolve the MAC address of
a packet's destination IP,  it is subject to ARP cache poisoning
attacks~\cite{volobuev1997playing}.

\subsubsection{Defenses}

There are several approaches to address ARP cache poisoning. Dynamic ARP
Inspection (DAI) \cite{cicscoarp} is a mechanism by which ARP responses are
checked against (1) a central DB that binds IP to MAC addresses (this may
be populated by listening to DHCP requests and responses in the network);
or (2) static pre-configured ARP entries.

Cryptographic measures can also be used to distribute pre-populated
IP-to-MAC-address mapping attestations, such as Ticket-based ARP (TARP)
\cite{lootah2007tarp}.  A voting-based protocol,  requiring network
consensus before updating ARP entries, has also been proposed by Nam \etal
\cite{nam2010enhanced}.

\subsection{Loop Free Forwarding}
\label{subsec:loopfree}

In networks with multiple routers and redundant physical paths, a routing
protocol is often used to advertise and learn routing information. Routing
protocols are either \emph{link state} (\eg IS-IS, OSPF---Open Shortest
Path First \cite{rfc2328}) or \emph{distance vector} (\eg RIP---Routing
Information Protocol \cite{rfc1058}).

\subsubsection{Attacks}

While attack methods vary among different routing protocols, a common attack
objective is \textit{routing table poisoning}---to pollute network topology
information and  derived forwarding tables by advertising or injecting
false routes through announcements. For example, a malicious router could
advertise a malicious Link State Advertisement (LSA) (\eg with a false
link cost) to influence other routers' calculation of routing tables. Such
attack is easy to launch but has limited impact since a neighboring router
will eventually advertise a correct LSA with a fresher sequence number,
resulting in the removal of the falsified LSA from being used for routing
table calculation. More advanced attacks (\cf ~\cite{Nakibly_BH2013}) can
be launched to increase the effectiveness of routing table poisoning.

\subsubsection{Defenses}
To mitigate routing table poisoning attacks, three levels of defenses should
be considered. First, routing protocols should only run in NNIs (routing
updates received from UNIs should be dropped). This is to prevent an outsider
(\eg a host) from participating in routing protocol communication. Second,
message origin authentication should be implemented to prevent a malicious
(compromised, previously legitimate) router from impersonating another
router. Third, routing updates should be corroborated when being used to
calculate routing tables.  For example, a link cost advertised by one router
should be corroborated with the link cost advertised by the other router on
the same link.

\subsection{Link Redundancy}
\lthree{} networks usually provide link redundancy through routing strategies
like Equal Cost Multiple Path (ECMP) routing~\cite{rfc2991,rfc2992}, which
allow packets to a common (\ie the same) destination address to be routed to
their next hops over multiple links of equal cost. Multiple path routing is
a local decision made within a single router, requiring no interaction with
adjacent or remote routers. Thus, it neither requires control protocols nor
appears to introduce new security risks.

\subsection{Device Redundancy}

If a gateway router goes down, traffic across different subnets will be
unable to reach their destinations, resulting in service outage. To improve
availability, two or more routers often share a common virtual IP address
and run a control protocol such as Virtual Router Redundancy Protocol
(VRRP)~\cite{rfc5798} to dynamically elect a master as the default gateway
of a subnet. When a master fails, VRRP dynamically selects another router as
the master. VRRP runs over IP with an IP multicast address as its destination.

\subsubsection{Attacks}

Protocols for high availability routing such as VRRP~\cite{rfc5798} are
subject to spoofing attacks. For example, VRRPv3~\cite{rfc5798} does not
include authentication of VRRP messages, thus an attacker may send a spoofed
VRRP message with the highest priority to become the master of the router
cluster. Such an attacker will receive all traffic to and from a subnet. While
the previous versions of VRRP~\cite{rfc2338,rfc3768} do include message
authentication, it was removed from version~3 because it could be exploited to
result in (malicious) election of multiple masters~\cite{rfc5798}. It is also
argued~\cite{rfc5798} that other attacks, such as ARP spoofing, exist which
could result in the same attack effect (\eg becoming the gateway of end hosts).

\subsubsection{Defenses}

Dropping VRRP messages that arrive from a host-connected port prevents
an attacker sitting at the network edge from spoofing such messages
\cite{rfc5798}. Additionally, VRRP message includes a TTL set to 255 by
default. Upon receiving a VRRP message, a router validates the TTL field and
discards a VRRP message whose TTL is not equal to 255. This limits the ability
of remote attackers (\eg outside of a network) from spoofing VRRP packets.

\subsection{Scalability}
Within an AS, scalability in \lthree{} networks is provided using routing
protocols, supported by hierarchical routing. For example, OSPF allows a
large network to be divided into sub-domains (OSPF areas). Routers within an
OSPF area need only maintain network topology information of the area they
belong to. A backbone OSPF area is used to connect all other areas. Thus
routing advertisements are limited to within an area, reducing the size of
routing databases. Between ASes, BGP is used to advertise network reachability
information.

Security risks of routing protocols, and their mitigation, are discussed in
Section~\ref{subsec:loopfree} above.

\section{Software Defined Networks}
\label{sec:sdn}

In SDN, the control plane of a device is implemented in an external entity,
as opposed to within the device in a \CN{}. This architectural difference
impacts how network properties from our framework are provided. More
specifically, in \CN{}s, a property achieved in the data plane is also
considered achieved in the control plane. This does not hold in SDN due
to the separation of planes. Thus for SDN, we discuss separately how each
network property is provided for the data and control planes. We use OpenFlow
switches~\cite{openflow2014} as an example in our discussion.

\subsection{Basic Forwarding}
Here we consider how a single SDN controller, controlling a single OpenFlow
switch connected with a number of hosts, learns forwarding information. We
assume that the controller has a direct connection with the switch, thus no
need to learn about this control connection.

 To configure the switch to provide connectivity among connected hosts, the
 controller must learn the mapping between hosts (\eg their MAC addresses)
 and switch ports. To do so, the controller may configure the switch to
 forward ARP requests and unknown packets, to the controller. An OpenFlow
 switch forwards such a packet to a controller using the \texttt{PACKET\_IN}
 message, which includes the switch port from which the packet is received. The
 \texttt{PACKET\_IN} message provides the controller with information about
 which hosts connect to which switch ports. If a destination host is also
 unknown, the controller instructs the switch to flood this packet using a
 \texttt{PACKET\_OUT} message. The response from the destination will also
 be sent to the controller, allowing the controller to learn the location
 of both hosts. As a result, the controller can set up flow rules for the
 pair to communicate.

This learning process by a controller, called \textit{Host Tracking Service},
is equivalent in principle to MAC learning by an \ltwo{} switch. It
demonstrates how a conventional \ltwo{} control function is taken out
from a switch, and implemented inside a controller. One difference is
that MAC learning, being \ltwo{}, only learns MAC addresses (possibly VLAN
IDs). OpenFlow can learn both \ltwo{} addresses (MAC address and possibly
VLAN ID) and IP addresses. We call this process \textit{Host Learning},
for better comparability with MAC learning.

\subsubsection{Attacks}
Host learning by a controller, being based on information provided by a switch
and hosts, is thus subject to spoofing attacks (MAC spoofing, IP spoofing,
VLAN tag spoofing), since a dishonest host or switch can forge such information
inside a packet. Often called {\it Host Location Hijacking} \cite{HONG_NDSS15},
here we call it {\it Host Profile Poisoning} for better comparability with
MAC table poisoning. Host learning is also subject to flooding attacks---an
attacker may generate a large number of packets with arbitrary MAC and IP
addresses, resulting in creating (1) a large number of host profiles inside
the controller, (2) a large number of messages sent to the controller, and
(3) a large number of flow tables inside a switch. Thus, it is possible to
cause DoS or packet interception, as unknown packets are also flooded. For
example, if the memory allocated for host profiles in a controller is full,
an existing host profile (\eg the oldest) will be overwritten, resulting in
the flooding of a new packet destined to that host. This resembles an \ltwo{}
MAC flooding attack.

Host-learning messages between switches and the controller may be exploited
to cause a message forwarding loop.
\subsubsection{Defenses}
 \textit{MAC binding} (discussed earlier) can mitigate MAC spoofing attacks,
 albeit requiring static configuration. MAC limiting (also discussed earlier)
 can mitigate MAC flooding attacks, but cannot prevent MAC spoofing.

To mitigate VLAN spoofing, an OpenFlow switch can designate its ports as
UNIs and NNIs, and remove VLAN tags in packets received from UNIs. Note since
there is only one switch, NNIs will not receive any traffic but the defense
still works in the case of multiple switches. If the port an SDN controller
connects to cannot be determined, and the controller needs to tag traffic
for some reason, this defense becomes problematic.

To mitigate IP spoofing, OpenFlow controllers could avoid the learning of IP
based forwarding rules from the data plane by planning IP address assignment
and configuring flow tables with IP prefixes, acknowledging that prefix
matching might be slower than precise matching.

\subsection{Loop Free Forwarding}
\begin{figure}[!ht]
 \centering
    \includegraphics[scale=0.4]{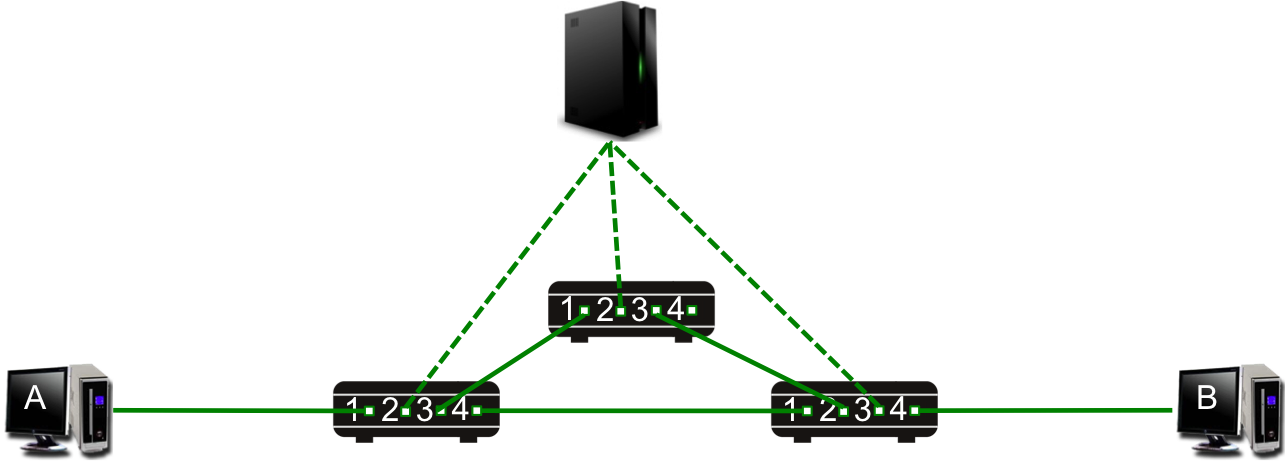}
  \caption{An SDN with multiple switches.}
  \label{fig:3switch-sdn}
\end{figure}

Here we consider a single controller controlling a number of OpenFlow switches
(Fig.~\ref{fig:3switch-sdn}).

To configure forwarding tables on OpenFlow switches, the controller must
first learn the network topology using a control protocol such as Link Layer
Discovery Protocol (LLDP)~\cite{ieee802station}. There are two scenarios
to consider. (1) There exists a dedicated control network (\eg a direct
link between each switch and the controller) such that each OpenFlow switch
can communicate with the controller, \eg by establishing a TLS connection
with the controller. (2) No such dedicated control network exists, and the
controller must discover all switches and set up proper flow tables so they
can begin communicating with the controller to receive further flows.

In case (1), the switch initiates communication with the controller during
a boot-up process. The controller thus obtains information about individual
switches under its control without running any control protocol. However,
the controller does not know the connectivity between switches, \ie the
network topology. In case (2), the controller does not know which switches
are under its control and must use a protocol to discover them (\cf Section
3.3 of ~\cite{cas_uss06}). Here we consider case (1), which is also commonly
studied in other papers (\eg~\cite{HONG_NDSS15}).

For topology discovery, an OpenFlow controller sends an LLDP packet (inside
a \texttt{PACKET\_OUT} message with output port set to ALL) to every switch
under its control. A switch receiving such a message floods to all of its
ports. A switch receiving an LLDP packet from a neighbor switch must forward
it to the controller, \eg due to the absence of flow rules for processing
such a packet, triggering a default rule for forwarding unknown packets to
the controller, or the existence of an explicit flow rule for forwarding
LLDP packets to the controller. This \texttt{PACKET\_IN} message to the
controller also includes the port number that receives the LLDP packet. Thus,
the controller discovers a link between two switches, and subsequently all
links between all switches, allowing completion of a complete network topology.

\subsubsection{Attacks}
An OpenFlow controller, discovering network topology by learning from
LLDP packets sent by switches, is thus subject to LLDP spoofing attacks. An
attacker, a switch or a host, may send falsified LLDP packets to a controller
to contaminate computation of network topology. For example, a host may send
falsified LLDP packets to insert itself  and create non-existent links into the
topology. This is a \textit{Link Fabrication Attack}~\cite{HONG_NDSS15}. LLDP
flooding may also force a controller to continuously re-calculate network
topology, disrupting service.

\subsubsection{Defenses}
To mitigate LLDP spoofing from hosts, an OpenFlow switch can designate its
ports as NNIs and UNIs, and reject LLDP packets received from UNIs. This
defense would work if the port an SDN controller connects to is prior
known. Otherwise, it will be problematic, since LLDP packets to and
from the controller might also be filtered. Message authenticity and
integrity, if implemented by LLDP, can effectively mitigate LLDP spoofing
by hosts. Consistency checks of LLDP packets by a controller can mitigate
LLDP spoofing by a host or switch. See Section~\ref{sec:comp-loopfree}
for discussion on how to mitigate attacks from an SDN controller.

\subsection{Link Redundancy}
We note two options for implementing link aggregation for two OpenFlow
switches with multiple physical links between them. (1) The OpenFlow switches,
just as an \ltwo{} switch, run LACP between them to create a virtual link
presented to the OpenFlow controller during link discovery. (2) An OpenFlow
controller discovers shared links between two switches and uses a group table
(available in OpenFlow 1.3 and later) to distribute traffic among a set of
switch ports within a group. Here the OpenFlow controller must monitor the
status of all links and update a corresponding group table upon detection
of a link state change. Note, if an OpenFlow switch shares multiple links
with a conventional switch or host running LACP, the OpenFlow switch can be
configured to pass LACPDUs from the conventional switch to the controller
to be processed. We consider the second case, since a switch implementing
LACP is considered hybrid rather than pure SDN switch.

\subsubsection{Attacks}
\label{sec:sdn-link-attacks}
An SDN controller is subject to message spoofing attacks. For example, an
attacker may send false link up and down events to a controller, to manipulate
the state of a link group. Such spoofing may be from a switch or host.

\subsubsection{Defenses}
Link state events should only be sent from OpenFlow switches, not hosts. If
the port an SDN controller connects to is prior known, such events if received
from other UNIs should be dropped. Message authenticity and integrity, if
implemented by OpenFlow events, can effectively mitigate spoofing by hosts
or switches. Consistency checks of OpenFlow events, if implemented by a
controller, can detect falsified events by a host or switch.

\subsection{Device Redundancy}

For redundancy among a group of OpenFlow switches, a controller could monitor
the status of each switch, and update switch configurations and flow tables
accordingly to allow traffic go through a new switch.   For redundancy among a
group of OpenFlow controllers, OpenFlow switches are configured with multiple
controllers. When the current controller goes down, a switch establishes a
connection with a next controller. A distributed election protocol can also
be used to elect a master controller, often for SDN applications. When the
master fails, another slave replica becomes the master.  The result of such
master election can also be synchronized into the switch's configuration
of controller preference to ensure all switches are also controlled by the
same master. For example, ONOS uses ZooKeeper \cite{hunt2010zookeeper},
a tool implementing an election protocol for distributed coordination and
election. OpenDaylight uses Akka and Raft \cite{184040} for master election.

Further, controller states must also remain synchronized among
controllers. For example, ONOS implements an eventual consistency
model~\cite{vogels2009eventually}, in which a background process updates
written objects in all replicas periodically.

\subsubsection{Attacks}

 Beside spoofing attacks against device redundancy (\cf
 Section~\ref{sec:sdn-link-attacks}), an election protocol used by controllers
 to achieve redundancy is subject to spoofing attacks. First, a non-controller
 entity (\eg a host) may join the election process to cause undesirable
 results. Second, a misbehaving controller may be able to manipulate the
 master election process (\eg by manually picking the smallest allowable
 time before candidacy election) to become the next master.

For example, Akka used by OpenDaylight for controller clustering employs no
default security mechanisms \cite{akkasecurity}, \eg for integrity protection
of inter-cluster messages. In ONOS, we note no security mechanisms are used
to ensure the integrity of update information. A timestamp may be forged,
and if replicas are not properly authenticated, an attacker may impersonate
one of them and manipulate stored objects.

\subsubsection{Defense}

If the ports SDN controllers connect to are prior known, election messages
if received from other UNIs can be be dropped. Otherwise,  message origin
authentication and integrity need to be implemented for the election protocol
and state replication, \eg using mutually authenticated TLS among controllers,
to prevent an outsider from participating in or tampering with the election
process. Additional mechanisms, \eg information corroboration, are also
needed to mitigate attacks by a misbehaving controller (\eg compromised by
an attacker) participating in election with forged information.

\subsection{Scalability}

Unlike \CN{}s, which run control protocols such as IS-IS to implement a
network fabric for scalability and other benefits, OpenFlow controllers can
configure flow rules in a way to improve scalability. For example, an OpenFlow
controller may configure: (1) encapsulation rules at ingress switches, (2)
forwarding rules based on outer headers at intermediate switches, and (3)
decapsulation rules at  egress switches. In this way, the controller creates
tunnels shared by many individual flows and the intermediate switches only
need to be configured with the rules related to tunnels, not individual flow,
significantly reducing the number of rules required. Edge switches (ingress
and egress) also only need to be configured with the rules relevant to the
end hosts connected to them. In this way, OpenFlow controllers may create
a network fabric without need of running additional control protocols.

Scalability in the SDN control plane can be provided by dividing a network into
areas or domains, each controlled by one or more controllers. Controllers could
be peer-to-peer or hierarchical. In a peer-to-peer model, area controllers
synchronize states among themselves so that each maintains a consistent global
view of the network. In a hierarchical model, lower level controllers maintain
a subset of the global view; only a top level controller has the global view.

Distributed controllers need to communicate with each other to exchange
reachability and state information. There is currently no standard defined
for inter controller communication. A distributed protocol is usually
needed. For example, BGP is suggested to be the message exchange protocol
among SDN controllers~\cite{yin2012}. ONOS relies on a distributed databases,
\eg Cassandra \cite{lakshman2010cassandra} and Distributed Hash Tables (DHTs),
for distributing network topology and state information among controllers.

\subsubsection{Attacks}

Vulnerabilities may arise from distributed communication in a peer-to-peer
model, or from controller-to-controller communication in a hierarchical
model. For example, if BGP would to be used to exchange information
among controllers, vulnerabilities in BGP could be exploited to attack
SDN. Vulnerabilities could also arise from the distributed database if it
would to be used for synchronization among controllers. For example, without
proper configuration, Cassandra may be vulnerable to query injection attacks
\cite{ron2015no}.

\subsubsection{Defense}

As with any other distributed protocol such as an election protocol,
communication among controllers, either peer-to-peer or hierarchical,
must provide data origin authentication and message integrity to prevent
outsiders from participating in or tampering with the communication. Further,
an additional mechanism such as information corroboration is needed to
mitigate misbehavior by legitimate controllers. Note, unlike in a simple
network where the ports to which controllers connect to can be prior known,
it is hard to define a communication boundary for distributed controllers
to prevent outsiders from participating in the election process.

\section{Comparison}
\label{sec:comparison}
 Table~\ref{tab:control-both} summarizes the control functions, attacks and
 defenses noted above. Here we compare security risks and defenses of \CN{}
 and SDN control planes in both the \edgeonly{} and \allelement{} threat
 models of Section~\ref{sec:framework}.
\subsection{Basic Forwarding}

MAC table poisoning (against a \CN{} switch) and host profile poisoning
(against an SDN controller), the two major threats  respectively, are similar
in nature but differ in details. For example, the attack vector of MAC table
poisoning is MAC address spoofing, while both MAC and IP addresses could be
spoofed in host profile poisoning. Since an \lthree{} router with manually
configured routing table does not learn forwarding information from the data
plane, it is not subject to IP spoofing attacks. Another subtle difference
is related to the size limit of the MAC table and memory allocated to host
profiles when flooding is employed. MAC table size could vary from a few
thousand to a few million entries, depending on the vendor and model of
a switch; an SDN controller usually has larger memory and is thus less
vulnerable to such flooding.

A \CN{} is also less vulnerable to DoS attack (than an SDN network)
because the SDN controller itself is a new attack surface, as is the
link between a switch and an SDN controller, which could become a new
bottleneck~\cite{shin2013attacking}.

Defenses for MAC poisoning and host profile poisoning are similar. For
example, port security could be used in a relatively static network to bind
switch ports with MAC addresses for both \CN{} and SDN. In a dynamic network
where MAC addresses might change often (\eg in a data center with server
virtualization), static binding is problematic for both paradigms.

We do not discuss the \allelement{} threat model here since this simple
network consists of only one switch and one controller. If the switch or
the controller is malicious, the network would be completely compromised.

\subsection{Loop Free Forwarding}
\label{sec:comp-loopfree}

Conventional \ltwo{} and \lthree{} networks use STP and routing protocols (such
as OSPF) for loop free forwarding. SDN uses LLDP for topology discovery.

In the \edgeonly{} threat model, an end host can attack both \CN{}s and
SDN, resulting in incorrect forwarding tables by exploiting protocols'
vulnerabilities.  While the impact from such attacks appears comparable,
attack techniques will differ since the protocols exploited differ.

\newcommand{\abltwo}{MAC table poisoning (MAC spoofing and MAC
flooding)~\cite{convery2002hacking}}
\newcommand{\alltwo}{BPDU spoofing, tampering and
flooding~\cite{attack_STP,kiravuo2013survey}}
\newcommand{\arltwo}{LACPDU spoofing~\cite{ieee_802_1AX_2014}}
\newcommand{\aaltwo}{Stacking spoofing}
\newcommand{\asltwoa}{VLAN hopping~\cite{VLAN_attacks}}
\newcommand{\asltwob}{Switch impersonation~\cite{VLAN_attacks}}
\newcommand{\asltwoc}{Routing advertisement spoofing}
\newcommand{\asltwo}{\asltwoa; \asltwob; \asltwoc}

\newcommand{\ablthree}{ARP table poisoning~\cite{volobuev1997playing}}
\newcommand{\allthree}{Routing advertisement spoofing~\cite{Nakibly_BH2013}}
\newcommand{\arlthree}{Link DoSing~\cite{rfc2991}}
\newcommand{\aalthree}{VRRP message spoofing~\cite{rfc5798}}
\newcommand{\aslthree}{Routing advertisement spoofing~\cite{Nakibly_BH2013}}

\newcommand{\ablsdn}{Host profile poisoning~\cite{HONG_NDSS15}}
\newcommand{\allsdn}{Link fabrication~\cite{HONG_NDSS15}}
\newcommand{\arlsdn}{Spoofed link-manipulation messages
(\eg~\cite{ieee_802_1AX_2014})}
\newcommand{\aalsdn}{Master election manipulation~\cite{akkasecurity}}
\newcommand{\aslsdn}{BGP attacks, distributed DB attacks}

\newcommand{\dbltwo}{Port Security~\cite{ciscoportsec}; MAC binding and
limiting}
\newcommand{\dlltwo}{Root Guard~\cite{ciscosecindepth}; BPDU filtering
(prevent a host from masquerading as a switch)}
\newcommand{\drltwo}{LACP source port authentication (data plane
implementation)}
\newcommand{\daltwo}{Run master election process on dedicated ports,
authenticate devices involved in the process}
\newcommand{\dsltwo}{VLAN filtering on UNIs, and disabling VLAN auto
negotiation}

\newcommand{\dblthree}{Dynamic ARP inspection (DAI)~\cite{cicscoarp};
Ticket-based ARP (TARP)~\cite{lootah2007tarp}; Voting-based
protocols~\cite{nam2010enhanced}}
\newcommand{\dllthree}{UNI filtering and consistency check of routing
advertisements}
\newcommand{\drlthree}{N/A}
\newcommand{\dalthree}{UNI filtering of VRRP messages, TTL checks}
\newcommand{\dslthree}{UNI filtering and consistency check of routing
advertisements}

\newcommand{\dblsdn}{MAC binding, host location validation~\cite{HONG_NDSS15}}
\newcommand{\dllsdn}{UNI filtering of LLDP packets~\cite{HONG_NDSS15},
Authentication of LLDP and OFDP packets}
\newcommand{\drlsdn}{UNI filtering of control messages, mutual authentication
of control channel}
\newcommand{\dalsdn}{Authenticity and integrity in master election}
\newcommand{\dslsdn}{Authenticity and integrity in communication among
SDN controllers}

\newcommand{\wida}{4.6cm}
\newcommand{\widb}{4.6cm}
\newcommand{\widc}{5.2cm}
\newcommand{\gapp}{13}
\newcommand{\linehere}{}

\begin{table*}[!ht]
\centering
\scalebox{0.9}{
\begin{tabular}{ r | L{\wida}  L{\widb} | L{\widc} c}\toprule[0.11em]

\multirow{2}{*}{\sc{\textbf{Network Properties}}} &
\multicolumn{2}{c|}{\sc{\textbf{Conventional Networks}}} &
\multicolumn{2}{c}{\multirow{2}{*}{\sc{\textbf{SDN}}}}	\\
& \multicolumn{1}{c}{\sc{\textbf{Layer-2}}} &
\multicolumn{1}{c|}{\sc{\textbf{Layer-3}}} &&\\

\bottomrule
\multicolumn{5}{c}{\cellcolor{gray!20}\it Control Functions}\\
\toprule

{\it Basic Forwarding}	& MAC Learning	& Static routes, Address Resolution
Protocol (ARP) & Host Location Learning       &\\[\gapp pt]\linehere
{\it Loop Free Forwarding}& Spanning Tree Protocol (STP) & Routing Protocols
(OSPF, RIP) & Link Layer Discovery Protocol (LLDP)   &\\[\gapp pt]\linehere
{\it Link Redundancy}		& Link Aggregation (LACP) & Equal Cost Multiple
Path (ECMP) & Controller			    &\\[\gapp pt]\linehere
{\it Device Redundancy}     & Switch Stacking	& Virtual Router Redundancy
Protocol (VRRP) & Election Protocol		    &\\[\gapp pt]\linehere
{\it Scalability}			& VLAN, Network fabric (TRILL, VxLAN)
& Routing Protocols (OSPF, IS-IS) & BGP, distributed DB		 &\\

\bottomrule
\multicolumn{5}{c}{\cellcolor{gray!20}\it Attacks}\\

\toprule

{\it Basic Forwarding}	    & \abltwo	& \ablthree	& \ablsdn
&\\[20pt]\linehere
{\it Loop Free Forwarding}	& \alltwo	& \allthree	& \allsdn
&\\[\gapp pt]\linehere
{\it Link Redundancy}			& \arltwo	& \arlthree
& \arlsdn	&\\[\gapp pt]\linehere
{\it Device Redundancy}		& \aaltwo	& \aalthree		&
\aalsdn       &\\[\gapp pt]\linehere
{\it Scalability}			& \asltwo   & \aslthree     &
\aslsdn   &\\

\bottomrule
\multicolumn{5}{c}{\cellcolor{gray!20}\it Defences}\\
\toprule

{\it Basic Forwarding}		& \dbltwo	& \dblthree	& \dblsdn
&\\[25pt]\linehere
{\it Loop Free Forwarding}	& \dlltwo	& \dllthree & \dllsdn
&\\[25pt]\linehere
{\it Link Redundancy}			& \drltwo	& \drlthree	&
\drlsdn       &\\[25pt]\linehere
{\it Device Redundancy}		& \daltwo	& \dalthree & \dalsdn
&\\[25pt]\linehere
{\it Scalability}				& \dsltwo	& \dslthree
& \dslsdn	&\\

\bottomrule[0.11em]

\end{tabular}
}
\caption{A summary comparison between Conventional Networks (\CN{}s) and SDN.}
\label{tab:control-both}
\end{table*}

Defenses for \CN{}s can rely on UNI filtering. However, that works in SDN
only if the attach points of the SDN controller are prior known and remain
static. Otherwise, UNI filtering is ineffective and crypto mechanisms are
required in SDN to prevent outsiders from participating in topology discovery.

In the \allelement{} threat model, the network (\eg \ltwo{} or \lthree{}
device and SDN controller itself) could be malicious; risks and defenses then
appear similar for \CN{}s and SDN, albeit with subtle differences. From a
risk perspective, a malicious \CN{} device or an SDN controller may be able
to compromise the entire network (\eg influencing the routing table of any
device within the network), acknowledging that a malicious SDN controller
appears capable of causing more damage.

From a defense perspective, data origin authentication, message integrity,
and consistency checks are all required by both \CN{}s and the SDN control
plane to counter insider attacks (\eg to detect and discard false information
received from other legitimate nodes). In CNs, consistency checks can be
done inside individual devices. In SDN, consistency checks should be done by
both switches and controller. First, an OpenFlow switch should validate LLDP
messages to ensure that it does not contain false information (\eg a faked
link between the sender and the receiver). Second, an SDN controller should
validate LLDP messages to rule out faked nodes and faked links. However, new
defenses are needed to mitigate or reduce risks of a controller misbehaving in
doing network topology discovery and route calculation. It appears difficult
to contain damage from a misbehaving controller if it is monolithic. Thus,
a controller is better divided into small, independent units to minimize
the risk from a misbehaving control unit and facilitate cross-checking the
behavior of each unit.

If adopted for SDN controllers, a micro service
architecture~\cite{DragoniGLMMMS16} can serve this purpose. As an example of
such an architecture, the control function providing loop free forwarding can
be implemented in three micro services, the first collecting and validating
LLDP messages, the second performing topology and route calculation, the
third updating flow rules in switches. Each micro service runs multiple
instances, each of which cross-checks requests and responses from multiple
other service instances. To cross-check behavior of flow rule updating
services, other types of services such as real time flow validation (\eg
VeriFlow~\cite{khu_hotsdn12}) can be implemented.

\subsection{Link Redundancy}

A \CN{} uses LACP for link aggregation; an SDN controller can monitor link
state changes, and update link groups in a switch accordingly. Both are
vulnerable to message spoofing and tampering attacks.

In the \edgeonly{} threat model, rules can be configured on UNIs to filter
LACPDU packets for \CN{}s. If the attach points of the SDN controller are
prior known and static, rules can also be configured to filter link up/down
events for SDN. Otherwise, crypto mechanisms are required by SDN to prevent
outsiders from sending link up/down events to the SDN controller.

In the \allelement{} threat model, message origin authentication and message
integrity can be used to address a legitimate switch spoofing a link group
member. If a link group member itself misbehaves, it falls short for the other
member to maintain a correct link group state since the misbehaving end can
manipulate packets (\eg selectively dropping them) to achieve the same end.

In SDN, message origin authentication and integrity, \eg by mutually
authenticated TLS, can mitigate a legitimate switch spoofing another switch by
sending the controller faked link up and down events. Further, the function
controlling link redundancy can be implemented in micro services, several
running simultaneously to cross-check each others' behavior.

\subsection{Device Redundancy}

Both \ltwo{} and \lthree{} use an election protocol to exchange messages
among a device group to elect a master, thus being subject to spoofing
attacks. In SDN, an election need not be implemented in switches for data
plane redundancy, but is required for controller redundancy.

In the \edgeonly{} threat model, rules can be configured on NNIs to filter
control messages from end hosts. If the attach points of the SDN controller
are prior known and static, such an approach can also be employed for
SDN. Otherwise SDN requires message origin authentication and message
integrity to counter outsider attacks.

In the \allelement{} threat model, message origin authentication and message
integrity are required to prevent one legitimate device from impersonating
another. An additional mechanism appears required to detect a legitimate
device from participating in an election using false information.  A similar
mechanism appears required in SDN.

\subsection{Scalability}

For scalability in \CN{} and SDN, respectively,
routing protocols such as OSPF and BGP can be used.  They are subject to
similar attacks. Regarding defenses, the network boundary can be defined for
\CN{} to discard control messages from end hosts in the \edgeonly{} threat
model. In SDN this is less effective since SDN controllers often run insider
servers that connect to the edge of a network; SDN controllers require crypto
mechanisms to prevent outsiders from participating in the routing protocols.

In the \allelement{} threat model, to detect inside attacks, \CN{}s and
SDNs require message origin authentication and message integrity, as well
as consistency checks.

$~~$

\noindent
\textit{Discussion.}
We save summary observations for Section~\ref{sec:discussion}.

\section{Related Work}
\label{sec:related}

To complement the preceding comparative analysis, we discuss relevant
literature for security in SDN and counterparts in conventional networks
(\CN{}s). Broad surveys of SDN security are available elsewhere (\eg
\cite{7150550,ahmadsecurity,alsmadi2015security}).

\subsection{Control Plane}

\subsubsection{Security-oriented controllers}

As one of the first network operating systems, NOX~\cite{GUDE_CCR08} provides
greater flexibility to the management plane. Despite lacking the ability
to undertake most network functionalities by itself, NOX aims to provide
sufficient APIs to ease the fulfilment of such functions. Porras \etal
\cite{POR_NDSS15} proposed SE-Floodlight, a security enhanced system based on
Floodlight \cite{flldlight}. Network administrators manually assign roles to
applications, while SE-Floodlight mediates all OpenFlow operations to enforce
a role-based permission model. SE-Floodlight also provides authentication
and flow conflict resolution services (based on FortNOX~\cite{POR12}),
both occurring on the system level independent of the applications, to
enforce privilege separation (\cf also PermOF~\cite{WEN13}). Shin \etal
\cite{shin_ccs14} identified several reasons for controller weaknesses,
including lack of (1) resource control, (2) application separation,
(3) application authentication and authorship, and (4) access control,
and presented Rosemary as a (non-monolithic) micro-NOS to address these
shortcomings.

\subsubsection{Control plane security extensions and APIs}

Many security extensions to SDN controllers have been proposed to monitor
and detect suspicious network behavior. VeriFlow \cite{khu_hotsdn12} and
FlowVisor \cite{SHE_OSDI10} are two examples. The former provides real time
checking and verification of forwarding behavior, whereas the latter enables
network slicing such that each slice is typically under a different control
domain, thus providing logical separation of multiple controller instances
(\cf~\cite{gutz2012splendid}). For such multi-slice networks, FlowChecker
\cite{als_safe10} is an example of a tool that checks consistency across
multiple slices.

In contrast to VeriFlow, FlowGuard~\cite{hu_hotsdn14} is a firewall for SDNs
that specifically focuses on conflict resolution. Note that SE-Floodlight
resolves conflicts only between flows, whereas FlowGuard resolves conflicting
network policies in general. To address the source-binding problem within the
network, FlowTags~\cite{fayazbakhsh2014enforcing} enables switches to tag
packets for appropriate source binding, avoiding conflict with middleboxes
in the network. Kim \etal \cite{Kim2015kinetic} proposed Kinetic to not only
monitor network properties, but also enable administrators to take appropriate
control actions in response to network changes, and to analyze source of errors
in control programs leading to the undesired network behavior. Similarly,
Flover~\cite{son2013model} and NetPlumber~\cite{kazemian2013real} are systems
to verify that flow policies do not contradict with desired network security
policies.

The proper extent of privileges that should be granted to an SDN
application is unclear. Excessive privileges may constitute a
significant weak point if an application becomes compromised or is
malicious~\cite{MAT_hotSDN14}; too few may not allow sufficient flexibility
to run security applications. Fresco~\cite{SHI13} is a framework for
developing security SDN applications, providing APIs for developers to
access sensitive network resources securely. Similarly, OperationCheckpoint
\cite{scott2014operationcheckpoint} aims to secure the network against third
party applications by ensuring that critical operations can be executed by
trusted applications only.

\subsubsection{Attack Mitigation}

Benton \etal \cite{BEN13} highlight the importance of isolating applications
running on top of the controller and the importance of verifying flow
tables, to avoid erroneous controllers (including errors introduced without
a malicious intent). Braga \etal \cite{braga2010lightweight} showed
how machine learning can be used to identify traffic involved in a DoS
attack. BASE~\cite{kwon2015incrementally} was proposed as an anti-spoofing
mechanism, aiming to mitigate DoS. SD-Anti-DDoS~\cite{cui2016sd} is another
tool used to \emph{clean} bloated flow tables after a DoS attack.

\subsubsection{Proposals for control plane scalability}

To ensure scalability and fault tolerance, numerous proposals
advocate replication and distribution of the control plane
\cite{botelho2014design,betge2015trust,heller2012controller}.
Onix~\cite{koponen2010onix} is a prominent example that abstracts
the network distribution state to the control plane running on
top. HyperFlow~\cite{tootoonchian2010hyperflow} allows multiple separate
SDN domains to be consolidated and controlled from a single point.

\subsection{Data Plane}

Despite SDN's promise to ease management and service deployment, it is becoming
clear that not all services can simply be implemented as applications on the
controller. Depending on required levels of network support, some security
solutions in the literature propose either low-level modules running in
controller kernel space, or software running on switches. These create
challenges in managing/updating network elements, as is the case with
\CN{}s. The OpenFlow Extension Framework (OFX)~\cite{sonchack2016ofx}
leverages control plane centralization to allow dynamic installation of
software on switches.

\subsubsection{Security services}

Distinct from \CN{}s, the centralization of the control plane in SDNs
challenges conventional means by which gateway services, \eg firewalls and
IDSs, are set up. For example, instructing edge switches to forward a copy
of the traffic to a separate IDS box would consume substantial delay and
bandwidth~\cite{mehdi2011revisiting}. FleXam~\cite{shirali2013efficient} is
an OpenFlow extension to enable a switch to send sample packets (including
payload) from a specific flow to the controller.
\subsubsection{Handling compromised switches}

Compromising an SDN switch enables a wide range of MitM and impersonation
attacks~\cite{antikainen2014spook}; detection in SDN  however differs
from \CN{}s. Chi \etal~\cite{chi2015detect} implemented applications that
periodically sample flow rules, and check if a random subset of switches
are behaving as instructed.

\subsubsection{DoS attacks}

While SDN is considered more vulnerable to DoS attacks~\cite{shin2013attacking}
than \CN{}s, mitigation techniques appear to follow non-conventional
solutions. For example, FloodGuard~\cite{wangfloodguard} rate-limits
packets sent to the controller to protect controller bandwidth from
being consumed by intentional \texttt{PACKET\_IN} requests from clueless
switches. SDNsec~\cite{sasaki2016sdnsec} allows edge switches to encode whole
routes on each ingress packet, thus mitigating DoS by state exhaustion. Flow
aggregation~\cite{kloti2013openflow} and flow time-out~\cite{kandoi2015denial}
are also widely regarded as good practices to mitigate switch memory
exhaustion.

\subsubsection{MitM attacks}

Sphinx \cite{DHA_NDSS15} aims to counter security threats from within an SDN
network (\eg from network switches and end hosts) by building flow graphs to
represent a closed form of the network topology, and using these to detect
anomalous switch behavior. Another tool, TopoGuard~\cite{HONG_NDSS15},
aims to detect LLDP hijacking and MitM attacks, by validating the network
view seen by the controller, to mitigate host location hijacking and link
fabrication vulnerabilities.

\section {Discussion and Concluding Remarks}
\label{sec:discussion}

Conventional networks and SDN share similar security risks related to the
control plane, but differ in defenses needed. We observe the following from
our analysis in Section~\ref{sec:comparison}. In the \edgeonly{} threat
model, conventional networks can define a network boundary to filter control
messages from end hosts. This approach is less effective in SDN---its control
plane is implemented in SDN controllers, which are usually connected to the
edge of the network; their attaching points, similar to end host locations,
may change unless there is a dedicated control network separate from user
networks. Thus, SDN largely requires cryptographic protection to prevent
outsiders from participating in the control plane.

In the \allelement{} threat model (Section~\ref{sec:trustmodel}), both
conventional networks and SDN require crypto mechanisms, as well as
consistency checks to mitigate insider attacks. They differ in where and
how such checks can be implemented. While straightforward consistency checks
might be implemented within individual \CN{} devices, it is less obvious where
and how to do this in SDN. Our analysis suggests that a highly modularized
and distributed SDN software architecture may facilitate consistency checks,
improving SDN control plane security (\cf~Section~\ref{sec:comp-loopfree},
including for modularization based on micro service architecture). As noted
earlier, current SDN controllers (\eg ONOS, Open DayLight) lack mechanisms
to handle insider threats.

Our framework for analysis of control plane security, suitable for both
conventional networks and SDN, is summarized by Table~\ref{tab:control-both}
and the discussion in Section~\ref{sec:comparison}. It allows
an apples-to-apples comparison, exploring both security risks and
defenses. An objective of our framework and analysis is to help guide
further SDN research, and to aid  practitioners in the design, development,
and deployment of software-defined networks with stronger robustness and
security properties. Future work includes designing SDN controllers that
can survive in the \allelement{} threat model, and means for comprehensive
testing of security vulnerabilities and defenses discussed herein.

\section*{Acknowledgement}

We thank Yapeng Wu and Xingjun Chu for their constructive discussion and
comments on this paper. The third author acknowledges funding from the Natural
Sciences and Engineering Research Council of Canada (NSERC) for both his Canada
Research Chair in Authentication and Computer Security, and a Discovery Grant.

\ifCLASSOPTIONcaptionsoff
  \newpage
\fi

\vfill

\end{document}